\documentstyle[preprint,aps,epsfig]{revtex}
\begin{document}

\tighten
\draft
\preprint{
\vbox{
\hbox{\today }
\hbox{ Tashkent}
}}
\renewcommand{\baselinestretch}{1.25}
\newcommand{\re}[1]{\ref{#1}}
\newcommand{\lab}[1]{\label{#1}}
\newcommand{\ci}[1]{\cite{#1}}
\newcommand{\bfr}{\begin{flushright}}
\newcommand{\bfl}{\begin{flushleft}}
\newcommand{\efl}{\end{flushleft}}
\newcommand{\efr}{\end{flushright}}
\newcommand{\bc}{\begin{center}}
\newcommand{\ec}{\end{center}}
\newcommand{\be}{\begin{equation}}
\newcommand{\ee}{\end{equation}}
\newcommand{\bea}{\begin{eqnarray}}
\newcommand{\eea}{\end{eqnarray}}
\newcommand{\ba}{\begin{array}}
\newcommand{\ea}{\end{array}}
\newcommand{\edc}{\end{document}}
\newcommand{\ul}{\underline}
\newcommand{\ri}{\rightarrow\infty}
\newcommand{\li}{\leftarrow\infty}
\newcommand{\ra}{\rightarrow}
\newcommand{\la}{\leftarrow}
\newcommand{\ds}{\displaystyle}
\newcommand{\dsf}{\displaystyle\frac}
\newcommand{\dt}{\Delta{t}}
\newcommand{\il}{\int\limits}
\newcommand{\pal}{\partial}
\newcommand{\xxx}{{\it{X}}}
\newcommand{\bone}{{\bf 1}}
%\sloppy

\title{
The simulation of the spin ground states of the
coulomb clusters in a broad  $2D$ parabolic well.      }

\author{
B. Abdullaev, A.D. Kidisyuk and M.M. Musakhanov
}
\address{ Institute of Applied Physics, National University of Uzbekistan,
Vuzgorodok, Tashkent, 700174, Uzbekistan}

\maketitle

\begin{abstract}

By variational Monte-Carlo method developed  Ceperley et al. for the
simulation of fermi systems in macroscopic confining potential well
we simulate various spin ground states of the coulomb clusters with
2,3 and 4 particles in a broad two-dimensional ($2D$) parabolic well.
In this method quantum state numbers determining the variational wave
function are not the numbers of well quantum states but numbers of the
equilibrium spatial positions of particles that give a minimum of the
system potential energy. The ground states with parallel, antiparallel
spins and as well, as bose state are simulated. For the cluster with
three particles it is also simulated the state when two particles have
one direction of spin and third opposite. The simulation shows that
clusters with parallel spins have lower ground state energy than clusters
with other spin configurations and bose state. That reminds a Hund's rule
in atomic physics when in not full filled atomic shells electrons prefer
to have a state with parallel spins.

\end{abstract}

\section{Introduction}

A considerable experimental succes in localization and cooling of
several ions in electrostatic traps \ci{th} and recent technological
advances in the fabrication of the quantum
dots in semiconductors \ci{k,ako} have induced
a big theoretical interest \ci{lm1,l,ldhb,j}
to these objects. A $2D$ parabolic potential well is a good model potential
well for the confining of electrons in the quantum dots \ci{bb} in many cases.
In the works \ci{lm1,l}  it was shown the equivalence of coulomb system in
effective electrostatic potential  $U_{ext}=aR^2$ to the Tomson model of
atom. The classic and quantum properties of coulomb clusters in traps
were also investigated in work \ci{lm3}.

Here we consider the coulomb clusters that correspond to quasi
two dimensional systems of ions or electrons taking  place in strong
anisotropic $2D$ traps that can be supposed as broad. By variational
Monte-Carlo method developed  Ceperley, Chester and Kalos \ci{cck} for
the simulation of fermi systems in macroscopic confining potential well
we simulate various spin ground states of the coulomb clusters with
2,3 and 4 particles in a broad $2D$ parabolic well.
In this method quantum state numbers determining the variational wave
function are not the numbers of well quantum states but numbers of equilibrium
spatial positions of particles that give a minimum of the system potential
energy. The ground states with parallel, antiparallel
spins and as well, as bose state are simulated. For the cluster with
three particles it is also simulated the state when two particles have
one direction of spin and third opposite. The simulation shows that
clusters with parallel spins have lower ground state  energy than clusters
with other spin configurations and bose state. That reminds a Hund's rule
in atomic physics when in not full filled atomic shells electrons prefer
to have a state with parallel spins.

\section{Model of system and the simulation method}

Let us consider a system - cluster having $N$ particles in a $2D$
parabolic well with Hamiltonian \ci{lm1}
\be
H=-\lambda\dsf{\pal^{ 2}}{\pal R^{ 2}}+U_{pot}(R),
\lab{1}
\ee
where $R$ corresponds to the two dimensional vectors
$(\vec r_1,\vec r_2,...,\vec r_N)$ and the potential energy is
\be
U_{pot}(R)=\ds\sum_{i=1}^N \vec r_i^{\ 2}+
\sum_{i>j}^N|\vec r_i-\vec r_j|^{-1}
\lab{2}
\ee

Hamiltonian (\re{1}) contains a single quantum parameter
$\lambda =\hbar^2a^{1/3}(2me^{8/3})^{-1}$,
where $m$ and $e$ mass and charge of coulomb particle
respectively, parameter $a$ determines a parabolic well
$U_{ext}=aR^2$ which  locates the particles. For a broad $2D$ parabolic
well we assume $\lambda<1$.

We simulate the coulomb clusters  by variational
Monte-Carlo method proposed in \ci{cck} for the simulation of fermi
systems in macroscopic confining potential well. It is based on the trial
variational wave function
\be
\Psi(R)=\psi_J(R)D,
\lab{4}
\ee
where to take into account an interparticle electrostatic correlation
and   a $2D$ parabolic well it is introduced the Jastrow
wave function:
\be
\psi_J(R)=\exp\left(-\beta\ds\sum_{i=1}^N\vec r_i^{\ 2}-\gamma\sum_{i>j}
|\vec r_i-\vec r_j|^{-1}\right).
\lab{5}
\ee
The expression for the quantity $D$ depends from the system simulating state.
If it is simulating a bose state of the system then
\be
D=\ds\prod_{i=1}^N\exp[-\alpha(\vec r_i-\vec r_{i0})^2].
\lab{6}
\ee
If there is a state of pure fermions then in according
with \ci{cck}
\be
D=\ds\prod_{s=1}^g\det(D_{ij}^s),
\lab{7}
\ee
where spin index $s$ has $g$ number of states (so for the parallel spin
state $g=1$ and for the antiparallel spin state $g=2$) and
$\det(D_{ij}^s)$ is Slater determinant, where
\be
D_{ij}^s=\exp[-\alpha(\vec r_i-\vec r_{j0})^2]_s.
\lab{8}
\ee
In the expressions (\re{6}) and (\ref{8}) the vectors $\vec r_{i0}$
provide a minimum of the potential energy
$U_{pot}(R)$ (\re{2}), i.e. a minimum of the potential energy for the
classical system. In the expression (\re{8}) index $i$ indicates the
particle's coordinate and $j$ indicates a quantum state.

The antiparallel spin state for the cluster with two particles
coincides with bose state of system.

The antiparallel spin state for the cluster with four particles
has a product of two Slater determinants. We assume that  in (\ref{8})
$j=1$ and $j=4$ describe an one direction of spin and
$j=2$ and $j=3$ the opposite direction. Therefore, in the expression (\re{7})
first determinant determines by $j=1,4$ and second by $j=2,3$.

The paralell spin state for the cluster with four particles  has only one
Slater determinant, where $j=1,2,3,4$.

A minimum of the ground state energy
\be
E=\dsf{\int dR \Psi^*(R)H\Psi(R)}
{\int dR |\Psi(R)|^2}
\lab{9}
\ee
is achieved by variation of parameters $\alpha$, $\beta$ and $\gamma$
in the expressions (\re{5})-(\re{8}).
It is assumed that  variational parameters $\beta$ and $\gamma$
in the Jastrow wave function (\re{5}) are the same for all spin and bose
states of all clusters (see \ci{cck}). That means that there is no
influence of the interparticle electrostatic correlation and external
field on the spin part of the wave function.

The cluster ground state simulation is occured by Metropolis algorithm.
In this algorithm each particle of the cluster is tested on the uniform
random displacement inside square with side $\Delta$  from center of
square $r_{j0}$. New spatial position of the particle is accepted
by the probability
\be
P=\min[1,|\psi(r_{new})/\psi(r)|^2].
\lab{10}
\ee
The size of $\Delta$ is chosen by the such way that the ratio of accepted
number to whole number of particle displacements has been roundly equal
$1/2$ for all particles. Thus, the full variation of energy (\ref{9}) is
taken place by the variation of four parameters $\alpha$, $\beta$, $\gamma$
and $\Delta$.

The mean values of the quantities in algorithm are calculated by the
formula:
\be
<F>=\dsf{\int dR \psi^*(R)F(R)\psi(R)}
{\int dR |\psi(R)|^2}\approx \dsf{1}{M} \sum_{i=1}^MF(R_i),
\lab{11}
\ee
where $R_i$ is the  $i$-th spatial configuration of the system
whole number of which is equal $M$.

The full energy of the system is the sum of every particle's energy.
Let's consider the expressions for the kinetic and potential energies,
for example, for the first particle.

The kinetic energy is
\be
E_{KIN}=-\dsf{\hbar^2}{2m}\int\Psi^*\Psi[(\vec\nabla_1\ln \Psi)^2+
\vec\nabla_1^2\ln \Psi]dR.
\lab{12}
\ee
If we introduce the determination
\bea\ba{c}
F_1^2=\left<\dsf{\hbar^2}{2m}(\vec\nabla_1\ln \Psi)^2 \right>,\\
T_1=\left<-\dsf{1}{4}\dsf{\hbar^2}{m}\vec\nabla_1^2\ln \Psi \right>
\lab{13}
\ea\eea
then the expression for the full energy will have a form:
\be
E_1=V_1+2T_1-F_1^2,
\lab{14}
\ee
where
\be
V_1 =\left<ar_1^2+\sum_{j=2}^N\dsf{e^2}{r_{1j}} \right>.
\lab{vr}
\ee
In the expressions (\re{13}) and (\re{vr}) the angle brackets mean the mean
value (\re{11}).

In the paper \ci{jf} it is shown that $T_1=F_1^2$ and therefore, we are
calculating the energy $E_1$ by the formula
\be
E_1=V_1+F_1^2.
\lab{15}
\ee

For the bosons containing cluster the expression for $F_1^2$ is
\bea\ba{c}
F_1^2=\dsf{\hbar^2}{2m}\left< \left|-\sum_{j=2}^N\vec\nabla_1u(r_{1j})-
\vec\nabla_1\chi(r_1)+
\vec\nabla_1\Phi_1(r_1)\right|^2 \right>.
\lab{16}
\ea\eea
Here $u(r_{1j})=\gamma/(r_j-r_1), \ \chi(r_1)=\beta r_1^2$ and
$$
\Phi_1(r_1)=\exp[-\alpha(\vec r_1-\vec r_{10})^2].
$$
If it is simulated the fermions containing cluster  then
\bea\ba{c}
F_1^2=\dsf{\hbar^2}{2m} \left< \left|-\sum_{j=2}^N\vec\nabla_1u(r_{1j})-\vec\nabla_1\chi(r_1)+
\sum_{j=1}^N\overline  D_{j1}^s\vec\nabla_1\Phi_j(r_1)\right|^2  \right>.
\lab{17}
\ea\eea
Here $\overline D_{ji}^s$ is the inverse matrix of matrix $D_{ij}^s$.

The main complexity of the simulation by the variational Monte-Carlo
method
of big number fermions containing systems is connected with technical
difficulty of inversion of matrix $D_{ij}^s$. In the paper \ci{cck} is
proposed a simple algorithm for the inversion of matrix $D_{ij}^s$
for these systems.
Our clusters have a small number of particles. Therefore, we use
direct  method of determination of elements of
inverse matrix $\overline D_{ji}^s$ by the elements of direct matrix
$D_{ij}^s$.

\section{The simulation results}

As it was said above, Hamiltonian (\re{1}) has a single quantum parameter
$\lambda =\hbar^2a^{1/3}(2me^{8/3})^{-1}$.
In accordance with this determination of parameter
$\lambda$ all length quantities are expressed in the
$e^{2/3}/a^{1/3}$ units and energy quantities in the  $e^{4/3}a^{1/3}$
units.

The positions of particles $r_0$ that give a minimum
of the potential energy  $U_{pot}(R)$ (\ref{2}) for the classic systems of
the clusters are determined analytically.
For all spin and bose states of one cluster they are suggested identical.

For the cluster  with two particles the positions of $r_0$
situate on the one line, on the distance $1/2$ from the
centre of coordinates that coincides with the centre of a parabolic well.

For the cluster with the three particles the coordinates of $r_0$
are the coordinates of the triangle
\bc$
r_{0x1}=-\left(\dsf{3}{2}\right)^{1/3}\dsf{1}{2}; \ \
r_{0x2}=\left(\dsf{3}{2}\right)^{1/3}\dsf{1}{2}; \ \
r_{0x3}=0;$ \\
$r_{0y1}=-\left(\dsf{3}{2}\right)^{1/3}\dsf{1}{2\sqrt{3}}; \ \
r_{0y2}=-\left(\dsf{3}{2}\right)^{1/3}\dsf{1}{2\sqrt{3}}; \ \
r_{0y3}=\left(\dsf{3}{2}\right)^{1/3}\dsf{1}{\sqrt{3}}.
$\ec

For the cluster  with four particles a minimum of $U_{pot}(R)$ provides by
square (with particles on the vertices of square and the centre of square
on the centre of a parabolic well and coordinates). The distance between
the centre  coordinates and square  vertex
is equal $r_0=\left(({4+\sqrt{2}})/({8\sqrt{2}})\right)^{1/3}$.

Variational parameters $\beta$ and $\gamma$ that are the same for all
spin and bose states of all clusters we
determine by the simulation of the bose state for the two particle
cluster. The numerical quantities of these are  $\beta=0.5$ and $\gamma=1.5$.

Whole number of displacements inside $\Delta$
for every particle we choose $10^4$, so absolute exactness of the
calculated mean energies is roughly equal $10^{-2}$ ( we would like to
note that for the Markov random processes, i.e. for independent random
processes, absolute exactness of the calculated in simulation mean
values is proportional to $1/\sqrt{N}$, where $N$ is number of random
simulations).

There is no a necessity for the simulation of the energies at every
numerical quantity of  $\lambda$ for every spin or bose state of the one
cluster. Let we have the calculated in the simulation the potential energy
$V$ and the kinetic energy $E_{KIN}$ at any one fixed $\lambda$.
As the kinetic energy in Hamiltonian (\re{1}) is directly proportional
to $\lambda$,
so for the getting of the kinetic energy for another $\lambda_1$\, one can
recalculate it by the formula $E_{KIN}(\lambda_1)=E_{KIN}(\lambda)\cdot
\lambda_1/\lambda$; the potential energy $V$ does not depend from $\lambda$
and one can be calculated once for this spin or bose state of cluster.

 Fig. 1 shows  variational parameter $\alpha$
dependence (in the  $e^{4/3}a^{1/3}$ units) of the potential energies
$V_{4BS}$, $V_{4AFS}$ and $V_{4FS}$ ( bose, antiparallel spin (antiferrospin)
 and parallel spin (ferrospin) states
respectively) for the cluster with four particles, of the
full energies
$E_{2BS}$ and $E_{2FS}$ ( bose and ferrospin states respectively)
for the cluster with two particles and of the kinetic energies
$E_{KIN4BS}$,
$E_{KIN4AFS}$ and $E_{KIN4FS}$  per one particle at the fixed
parameter $\lambda=0.01$.

In this figure the  potential energy for the bose state $V_{4BS}$ is
increased
when parameter $\alpha$  goes to zero. This takes place due the increasing
of the electrostatic interaction between coulomb particles
because the decreasing of $\alpha$ means the increasing of the mean size
of the wave function per one particle.
The $\alpha$ dependence of the potential energy for parallel spin state
$V_{4FS}$ when $\alpha$ goes to zero has opposite character. This is occured
due the increasing of the exchange interaction between fermions.

 Fig. 2 shows the $\lambda$ dependence (in the  $e^{4/3}a^{1/3}$ units) of
full ground state energy for every cluster per one particle. In this
figure the symbol $3AFS$ corresponds to the state when two particles
have one direction of spin and third opposite.
The numerical quantities of the full energies in Fig. 2 are
the minimums of these energies on  variational
parameter $\alpha$.

At last, for the demonstration of qualitative behaviour of the module of
the wave function,
we have simulated it for the cluster with three particles at
parameters $\beta=1.7$ and $\gamma=1.0$ (though these parameters of
$\beta$ and $\gamma$ do not give the minimum of the full energy, but they
provide for the module of the wave function
an obvious visual illustration). This quantity is outlined in the figures
3,4 and 5 for the bose,
parallel spin and mixed spin ( i.e. $3AFS$) states  respectively.
From these figures one can see clearly that exchange
interaction between fermions moves apart a quantum spatial distribution
of the cluster particles. Bose and fermi like together behaviours
of the module of the wave function are outlined in Fig. 5.

The main result of the present work is outlined in the figures  1 and 2.
It turns out that for the every cluster with 2,3 and 4 coulomb particles in
a $2D$ parabolic well the ground states with parallel spins are energetically
advantageous  than other spin and bose states.
That reminds a Hund's rule in atomic physics when by taking account of
exchange interaction in not full filled atomic shell it is profitable for the
electrons to have a single direction of spin.

We are grateful Yu.E. Lozovik for indication on this task and helpful
discussions.

\newpage

%%%%%%%%%%%%%%%%%%%%%%%%%%%%%%%%%%%%%%%%%%%%%%%%%%%%%%%%%%%%%%
\begin{figure}
\begin{center}
\parbox{10cm}{\epsfxsize=10.cm \epsfysize=10.cm \epsfbox[5 5 500 500]
{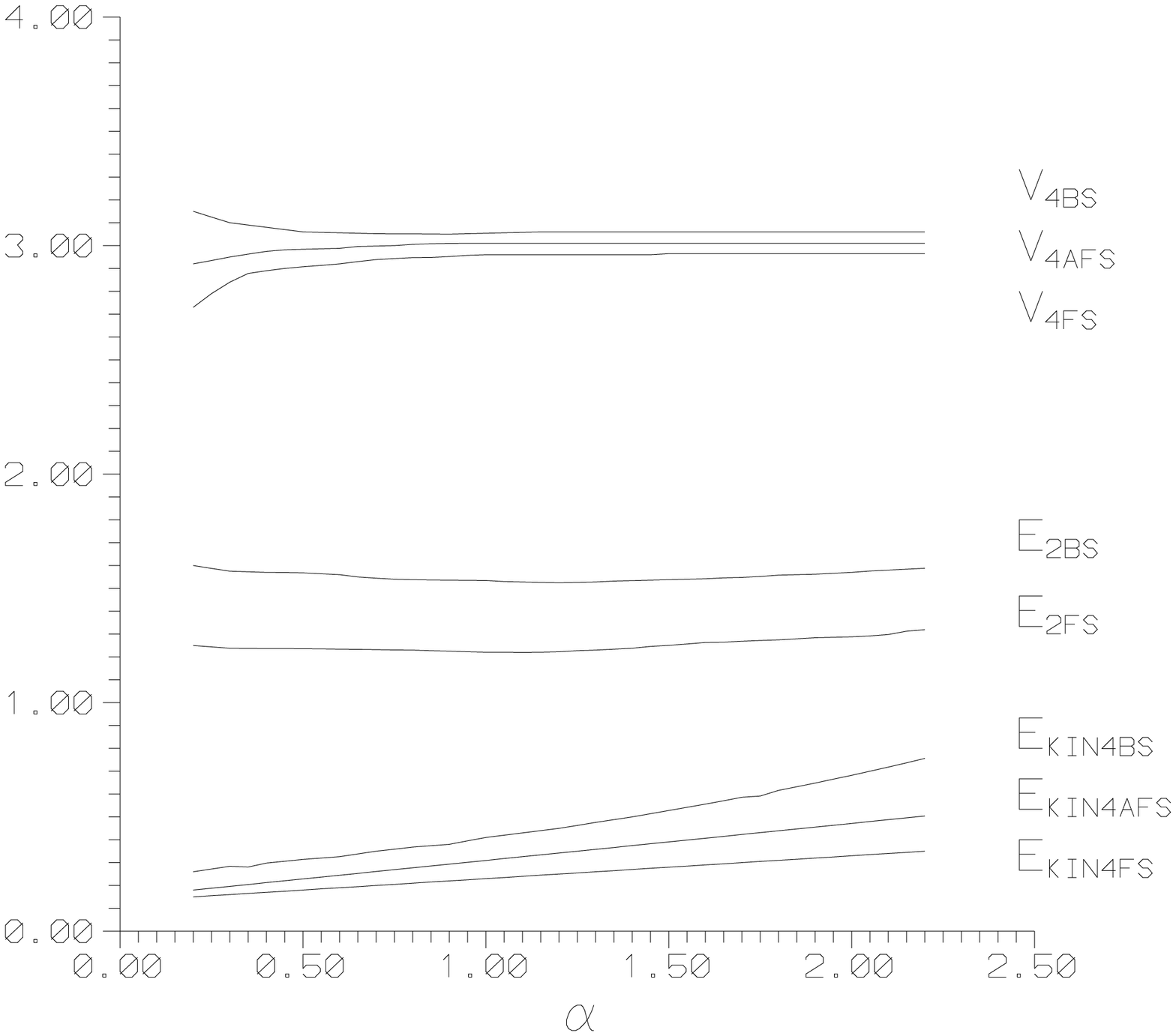}{}}
\end{center}
\caption{ The $\alpha$ dependence of the potential energies
 $V_{4BS}$, $V_{4AFS}$, $V_{4FS}$, the full energies
 $E_{2BS}$, $E_{2FS}$ and the kinetic energies $E_{KIN4BS}$, $E_{KIN4AFS}$,
 $E_{KIN4FS}$ at the fixed  $\lambda=0.01$.
 }

\end{figure}
%%%%%%%%%%%%%%%%%%%%%%%%%%%%%%%%%%%%%%%%%%%%%%%%%%%%%%%%%%%%%%
\begin{figure}
\begin{center}
\parbox{10cm}{\epsfxsize=10.cm \epsfysize=10.cm \epsfbox[5 5 500 500]
{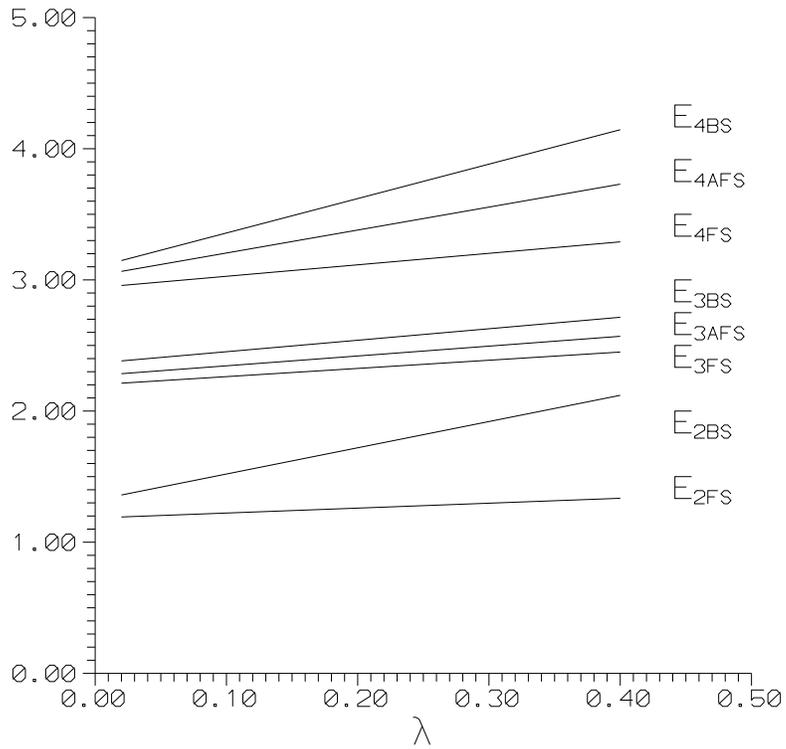}{}}
\end{center}
\caption{  The $\lambda$ dependence of the full energies for all clusters.
% $E_{4BS}$, $E_{4AFS}$, $E_{4FS}$, $E_{3BS}$,
%  $E_{3AFS}$, $E_{3FS}$, $E_{2BS}$,
% $E_{2FS}$.
 }

\end{figure}

%%%%%%%%%%%%%%%%%%%%%%%%%%%%%%%%%%%%%%%%%%%%%%%%%%%%%%%%%%%%%%
\begin{figure}
\begin{center}
\parbox{10cm}{\epsfxsize=10.cm \epsfysize=10.cm \epsfbox[5 5 500 500]
{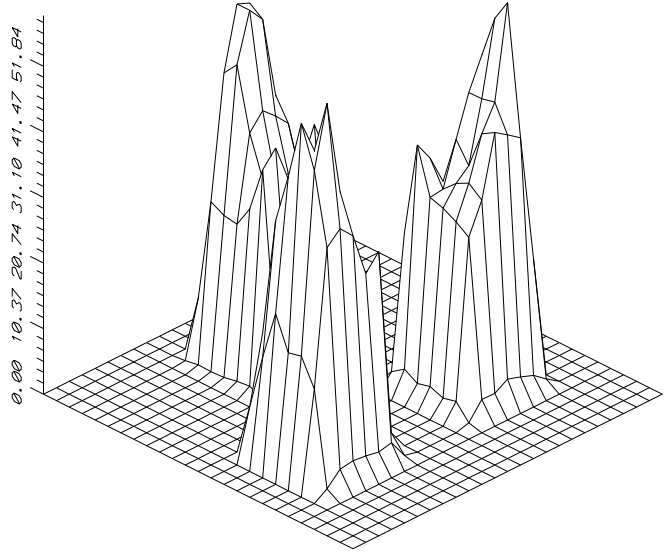}{}}
\end{center}
\caption{ The module of the wave function for the bose state
 for cluster with three coulomb  particles
      in a $2D$ parabolic well.}

\end{figure}

%%%%%%%%%%%%%%%%%%%%%%%%%%%%%%%%%%%%%%%%%%%%%%%%%%%%%%%%%%%%%
\begin{figure}
\begin{center}
\parbox{10cm}{\epsfxsize=10.cm \epsfysize=10.cm \epsfbox[5 5 500 500]
{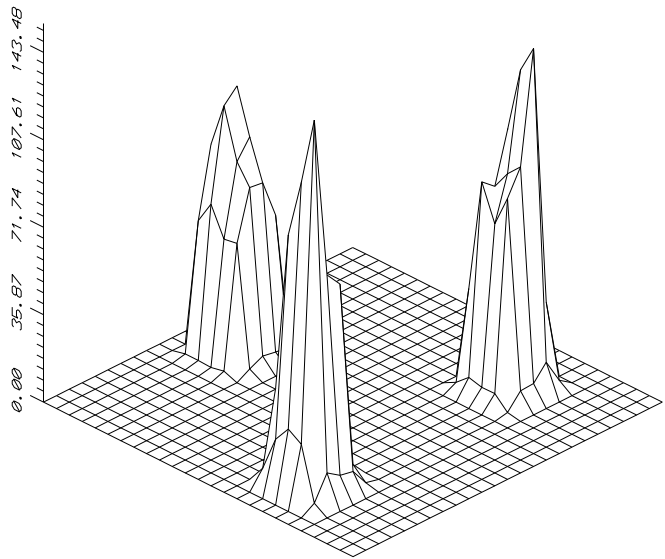}{}}
\end{center}
\caption{The module of the wave function for the parallel spin state
 for cluster with three coulomb  particles
      in a $2D$ parabolic well.}

\end{figure}

%%%%%%%%%%%%%%%%%%%%%%%%%%%%%%%%%%%%%%%%%%%%%%%%%%%%%%%%%%%%%
\begin{figure}
\begin{center}
\parbox{10cm}{\epsfxsize=10.cm \epsfysize=10.cm \epsfbox[5 5 500 500]
{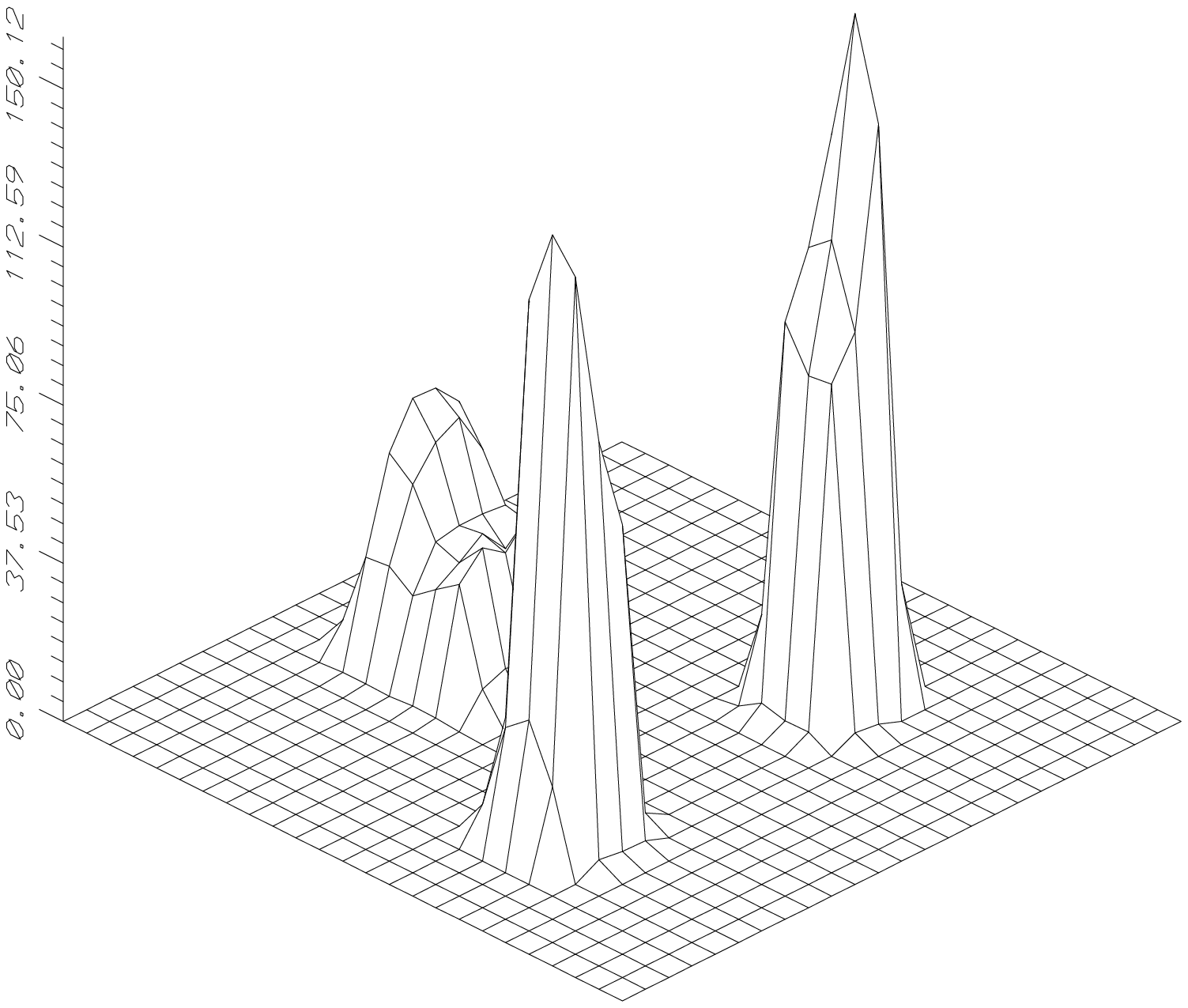}{}}
\end{center}
\caption{ The module of the wave function for the
mixed spin state for cluster with three coulomb  particles
in a $2D$ parabolic well when two particles have one
      direction of spin and third opposite. In the
      figures 3,4 and 5 the numerical quantities in the vertical axes expressed
      in single numerical units. }
\end{figure}

%%%%%%%%%%%%%%%%%%%%%%%%%%%%%%%%%%%%%%%%%%%%%%%%%%%%%%%%%%%%%

\edc
\begin{thebibliography}{99}


\bibitem{th}
P.E. Toschek, in Les. Houches, Session 38, New trends in atomic physics,
Vol. 1, eds. G.Grynberg and R.Stora (North - Holland, Amsterdam)
p.383, 1984; S. Becker et al., Rev. Sci. Instr., {\bf  66}, 4902 (1995);
H.J. Kluge, Nucl. Instr. Methods Phys. Res., {\bf B98}, 500 (1995).

\bibitem{k}
M.A. Kastner, Rev. Mod. Phys., {\bf 64}, 849 (1992); Phys. Today, {\bf 46},
24 (1993).

\bibitem{ako}
R.C. Ashoori, Nature (London), {\bf 379}, 413 (1996); L.P. Kouwenhoven,
T.H. Oosterkamp et al., Science, {\bf 278}, 1788 (1997).

\bibitem{lm1}
Yu.E. Lozovik and V.A. Mandelshtam, Phys. Lett., {\bf A145}, 269 (1990).

\bibitem{l}
Yu.E. Lozovik, Usp.Fiz.Nauk, {\bf 153}, 356 (1987);

\bibitem{ldhb}
B.C. Levi, Phys. Today, {\bf 21}, 17 (1988);
D.H. E. Dubin and T.M. O'Neil, Phys. Today, {\bf 60},  511 (1988);
J. Hoffnagle, R.G. de Voe, L. Reyna and R.G. Brewer,
Phys. Today, {\bf 61}, 255 (1989);
R. Blumel, J.M. Chen, E. Peik, W. Quint, W. Schleich, Y.R. Shen and
H. Walter,  Nature (London), {\bf 334}, 309 (1988).

\bibitem{j}
N.J. Johnson, J.Phys.: Condens. Matter, {\bf 7}, 965 (1995).

\bibitem{bb}
P. Bakshi, D.A. Broido and K. Kempa, Phys. Rev.,{\bf B42}, 7416 (1990);
N.A. Bruce and P.A. Maksym, Phys. Rev., {\bf B61}, 4718 (2000).


\bibitem{lm3}
Yu.E. Lozovik and V.A. Mandelshtam, Phys. Lett., {\bf A165}, 469 (1992).

\bibitem{cck}
D. Ceperley, G.V. Chester and M.H. Kalos, Phys. Rev., {\bf B16}, 3081
(1977).

\bibitem{jf}
H.W. Jackson and E. Feenberg, Ann. Phys. (NY), {\bf 15}, 266 (1961).

\end{thebibliography}
